\documentclass[namedreferences]{solarphysics}
\usepackage[hyperref,optionalrh]{spr-sola-addons}
\usepackage{natbib,graphicx,multirow,color,breakurl,enumitem}
\usepackage[margin=2.5cm]{geometry}

\newcommand{\aap}{    {\it Astron. Astrophys.}}

\newcommand{\apj}{    {\it Astrophys. J.}}

\newcommand{\mnras}{  {\it Mon. Not. Roy. Astron. Soc.}}

\newcommand{\solphys}{{\it Solar Phys.}}
 
\newcommand{\ssr}{    {\it Space Sci. Rev.}} 
\chardef\us=`\_

\begin{document}

\begin{article}

\begin{opening}
\title{Fast Solar Image Classification Using Deep Learning and its Importance for Automation in Solar Physics}

\author[addressref={aff1},corref,email={j.armstrong.2@research.gla.ac.uk}]{\inits{J.A.}\fnm{John A.}~\lnm{Armstrong} \orcid{0000-0003-1589-9365}}
\author[addressref={aff1,aff2}]{\inits{L.}\fnm{Lyndsay}~\lnm{Fletcher} \orcid{0000-0001-9315-7899}}

\address[id=aff1]{SUPA School of Physics and Astronomy, University of Glasgow, Glasgow, G12 8QQ, Scotland, U.K.}
\address[id=aff2]{Rosseland Centre for Solar Physics, University of Oslo, P.O.Box 1029 Blindern, NO-0315 Oslo, Norway}

\runningauthor{J.A.~Armstrong and L.~Fletcher}
\runningtitle{Fast Solar Image Classification}

\begin{abstract} 
The volume of data being collected in solar physics has exponentially increased over the past decade and with the introduction of the \textit{Daniel K. Inouye Solar Telescope} (DKIST) we will be entering the age of petabyte solar data.
Automated feature detection will be an invaluable tool for post-processing of solar images to create catalogues of data ready for researchers to use.
We propose a deep learning model to accomplish this; a deep convolutional neural network is adept at feature extraction and processing images quickly.
We train our network using data from \textit{Hinode/Solar Optical Telescope} (SOT) H$\alpha$ images of a small subset of solar features with different geometries: filaments, prominences, flare ribbons, sunspots and the quiet Sun (\textit{i.e.} the absence of any of the other four features).
We achieve near perfect performance on classifying unseen images from SOT ($\approx$99.9\%) in 4.66 seconds.
We also for the first time explore transfer learning in a solar context.
Transfer learning uses pre-trained deep neural networks to help train new deep learning models \textit{i.e.} it teaches a new model.
We show that our network is robust to changes in resolution by degrading images from SOT resolution ($\approx$0.33$^{\prime \prime}$ at $\lambda$=6563\AA{}) to \textit{Solar Dynamics Observatory/Atmospheric Imaging Assembly} (SDO/AIA) resolution ($\approx$1.2$^{\prime \prime}$) without a change in performance of our network.
However, we also observe where the network fails to generalise to sunspots from SDO/AIA bands 1600/1700\AA{} due to small-scale brightenings around the sunspots and prominences in SDO/AIA 304\AA{} due to coronal emission.

\end{abstract}

\end{opening}

\section{Introduction} \label{S-intro}
With each new solar physics mission/telescope, instruments are improving in spatial, temporal and/or wavelength resolution.
Increased resolution in any of these three categories equals greater volumes of data.
This has led to an exponential increase in the amount of data acquired in the past decade, from \textless~10TB per year from \textit{Hinode/Solar Optical Telescope} (SOT) in 2006 \citep{2008Tsuneta} to 500TB per year from the \textit{Solar Dynamics Observatory} (SDO) in 2012 \citep{2002Schwer} to 10000TB per year expected from the \textit{Daniel K. Inouye Solar Telescope} (DKIST) which is seeing first light in 2019 \citep{2014Elmore}.
On top of this, the \textit{Hinode} and SDO data is all archived\footnote{\textit{Hinode} data available from \url{http://sdc.uio.no/sdc/} and SDO data available through VSO in SolarSoft and SunPy.} totaling 3.1PB (petabytes) of data which will only keep growing with each passing year.

This is a huge amount of data, and sorting through it is not a task which can be given to humans.
For an efficient alternative, we must turn to automation, particularly machine learning.
Machine learning is the process of using statistical techniques to give computers the ability to learn how to perform a certain task without being explicitly programmed.
In the past two years, applications of machine learning techniques in solar physics have seen a rise in popularity, being applied to complex problems such as the inversion of solar flare atmospheres \citep{2019radynversion}; magnetogram super-resolution \citep{2018DiazBaso}; photospheric horizontal velocity field calculations \citep{2017AsensioRamos}; real-time multi-frame solar image deconvolution \citep{2018AsensioRamos}; solar flare and space weather forecasting \citep[][to name a few]{2018Florios,2017Liu,2018Liu,2017Nishizuka,2018Piana}; looking at the spatial and temporal correlations of spectral profiles during flares \citep{2018Panos} and detecting emergent flux \citep{2018hao}.
Machine learning in solar physics has proved its worth and will only continue to increase in usage as time goes on.
On top of performing complex tasks, machine learning is also adept at data management and dataset reduction.
This is the kind of automation that can save data analysts time and effort when acquiring and traversing their data and in the age of data-intensive solar physics these techniques can prove invaluable.

Motivated by this, we propose an efficient machine learning algorithm for the classification of solar images: a convolutional neural network (CNN).
This is designed to learn the different geometry of large-scale features on the Sun such that, after the model has been trained, a dataset of solar images can be passed to the network and it will identify which images contain which relevant feature in a very short time.
This process has already been used for galaxy classification in cosmology \citep{2018Dai,2018Alhassan} and we propose adopting a similar algorithm for solar purposes.

The algorithm we propose will allow the user to easily identify the images of most importance to the study they are carrying out.
Furthermore, having a pre-trained CNN that understands the geometry of solar features can be very beneficial for ``transfer learning''.
Transfer learning is when a previously trained neural network is used for initialisation and/or training for a new network which aims to learn a different but related task.
Neural networks themselves approximate a functional representation of a process from the input to the output that is learned through a complex optimisation problem known as training.
This is known as the Universal Function Approximation Theorem and was proven for fully-connected layers with sigmoid activations by \citet{1989Cybenko} and more generally with rectified linear unit (ReLU) activations by \citet{2017Lu}.
Training is performed on a high-dimensional space and thus contains many local minima which can correspond to non-physical solutions, and optimisers can get stuck there.
This is the most difficult problem to overcome when applying neural networks to physical data.
Therefore, transfer learning can be beneficial as the old network teaches the new network what it knows about the physical system and can steer the optimiser towards a physical solution.

We are interested in optical wavelengths and will focus on images within this range.
We train our CNN using images from \textit{Hinode}/SOT instrument taken by the H$\alpha$ (6563\AA{}) filter.
The images are sorted into 5 classes: filaments, flare ribbons, prominences, sunspots and the quiet Sun (\textit{i.e.} lack of any of the other four features).
Thus, the network learns the geometry of these features when observed at this wavelength.
One of our goals is to see if the computer perceptually understands what these features are.
That is, if it can identify the same features correctly when they are imaged in different wavelengths \textit{e.g.} sunspots observed in 1600/1700\AA{} and prominences observed in 304\AA{}.

The training set itself, is a catalogue of 13175 H$\alpha$ images from SOT that were classified by hand into one of the five classes.
The aim of this dataset was to take as wide a space of each feature as possible \textit{i.e.} the sunspot class contains images with single sunspots, multiple sunspots, different shapes and sizes of sunspots etc.
The catalogue of data we have amassed may be useful for anyone looking to train a machine learning model on solar image data or transferring pretrained knowledge of features on the Sun to their project\footnote{This can be downloaded from \url{http://github.com/rhero12/Slic/releases/tag/1.1.1}.}.
Furthermore, our data catalogue provides a good template for how to organise a dataset for a solar classification problem and may be beneficial for classification in other wavelengths such as radio using LOFAR \citep{2004Bastian} or X-ray using RHESSI \citep{2000Lin}.

The structure of the paper is as follows: in Section~\ref{S-cnn} we give a comprehensive overview of deep learning and how a convolutional neural network actually learns a task. In Section~\ref{S-model}, we describe our model architecture and the specifics of how we trained. In Section~\ref{S-results}, we discuss statistical tests to challenge our model and the issue of expanding our network beyond visible wavelengths. Finally, in Section~\ref{S-discussion}, we present our conclusions and examples of how the model we have developed can be used beyond the scope of this paper.

\section{Convolutional Neural Networks} \label{S-cnn}
\subsection{Background} \label{S-bg}
Neural networks (NNs) are very powerful universal function approximators meaning that they have the (theoretical) ability to model any unambiguous function no matter how complex.
To understand NNs, we must look back to the conception of machine learning, namely Rosenblatt's perceptron.
This is a simple setup modelled on a neuron in the brain: there are many inputs with varying electrical signals which are integrated, and depending on a threshold the neuron will either fire or not.
This translates to there being an input dataset and a weight vector (which are the parameters learned by the system).
The vector inner product is found between the input and the weights before being passed to a step function which decides whether or not the neuron fires.
This is the basis for nodes in NNs (however modern nodes are more generalised in the linear and non-linear part of the transformation).
Each node has a different associated weight vector and a linear transformation. Then non-linearity is applied to the input which will determine the nature of the signal outputted by each node.
An NN is a system of interconnected nodes which learn to perform a specific task after being trained in a supervised manner.

Supervised learning means that the data used for training has a defined structure.
That is, the user knows what the function between the input and output should produce in the prediction stage (be it classification or regression).
In supervised learning, we want the algorithm to learn the functional approximation by understanding the pre-defined structure between the input and output.

The simplest NN example is that of a shallow neural network (SNN) -- this is a neural network that consists of one hidden layer, where a hidden layer is a layer of nodes that comes between the input and the output.
This, however, can only learn simple functions in a reasonable amount of time.
To learn more complex functions, we must look towards ``deep'' learning.
Deep learning is the practice of stacking more than one intermediate layer in an NN.
The introduction of more layers allows the network to learn more complex functions in a smaller amount of time.
While it is true that a single hidden layer can learn any non-linear function \citep{1990Jones}, it will need increasing amounts of time to do so, such that it is more efficient to stack hidden layers as each hidden layer will learn a part of the overall non-linear function \citep{1991Hornik}.

The connections made between different layers in an NN is important for the number of parameters that need to be learned and how efficiently and well this can be done.
Classic feed-forward neural networks (NNs with a linear graph) make use of fully-connected layers, in which each node in a previous layer is connected to each node in the next layer.
This utilises a linear transformation of the data followed by a non-linearity and will result in a very dense, high-dimensional matrix since every connection will have a corresponding weight vector filled with parameters to be learned.
This is an issue when it comes to image data as each individual pixel would need to be represented by a node.
For example, for a megapixel image, the layers in a fully-connected network (FCN) would have a width ~$\mathcal{O}$(10$^{6}$).
This means that there would be 10$^{6} \times$ \textit{N} calculations to be done between the input layer and the first hidden layer with width \textit{N}.
This is impractically computationally expensive.
To counter this, \citet{1998Lecun} introduced a concept that would revolutionise using machine learning for image-based tasks: convolutional neural networks.

\subsection{Convolutional Neural Networks} \label{S-cnn2}
Following the biological trend, CNNs are modelled after the visual cortices of animals.
The idea is that the visual cortex of an animal is a system of interconnected neurons, starting at the eye with an image and ending at the brain with an understanding of what is in the image, and passing a specific electrical signal between the connected neurons depending on the features that each layer identifies.
This is achieved in a hierarchical manner meaning that the first layers identify low-level features (\textit{e.g.} colours, gradients, lighting) followed by later layers identifying high-level features (\textit{e.g.} facial features).
This means that the biological neurons do not detect specific features of objects but rather each layer identifies an abstract feature whose signal will help the subsequent layers to pick out other abstract features and the final combination of these abstract features tells the brain what the animal is seeing.
The animal in question then subconsciously teaches its neurons how to react to different objects \textit{i.e.} it learns.
This is what we want to accomplish with a CNN.
We set up our artificial network to model the biological network described previously with each deeper layer learning how to react to more and more abstract features in the image data such that the output layer will learn different objects' geometry.
This, however, comes at a cost of interpretability as the abstract features that the network chooses to learn in each image classification task are chosen by the computer itself.
In essence, the features are learned within the machine's imagination rather than being something we can explicitly write down.

Rather than having every node connected to every other node in the subsequent layer, convolution nodes make use of what is known as the receptive field: this is where the image is convolved with some kernel of a pre-defined size at certain locations on the image to produce what is known as a feature map.
This means that the linear transformation in each node is represented by the convolution function, as CNNs were specifically designed to take an image as input and it works well when making the analogy to the biological network described above.
The key ideas are that the convolution layer will extract features from the image and the network will learn to extract the most important features for the task at hand; and the combination of many feature detector layers will form a good understanding of the spatial arrangement of the pixels in an input image.
Each convolutional node produces a feature map, with each convolutional layer producing many, such that many features can be extracted and analysed at the same time.
This concept relies on the fact that neighbouring pixels in an image are strongly correlated with the correlation decreasing with distance from the pixels in question.
The other benefit to this is that, rather than there being a weight to train between every node, the convolutional kernel for each convolutional node  (\textit{i.e.} for each feature map) will be the trainable weights in the system (known as ``weight sharing'').
So, for a 1 megapixel image, rather than having 10$^{6}$ nodes with 10$^{6} \times$ \textit{N} weights, if a convolutional kernel of 3$\times$3 pixels is used in a convolutional layer producing 64 feature maps then there will only be 576 trainable parameters which is a huge computational advantage.

Deep CNNs are well-established in image classification tasks.
This is due to CNNs automatically dealing with any shift invariance \textit{via} the convolution function.
What this means is that CNNs learn the geometry of the features in the images rather than relative positions of pixels in images, which is important as features are not always going to be in the same spatial position or orientation in every image or be the same size in every image.
This leads to easier identification of objects in images as the network looks for these feature maps as opposed to specific sequences of pixels \citep{2003Simard}.

\subsection{Training} \label{S-training}
Training is the most important and difficult process in any machine learning algorithm.
Training is how a network learns what method it is supposed to be approximating.
In our case, we are employing supervised learning which described in Section~\ref{S-bg}.
This is implemented in the network \textit{via} a feedforward and backpropagation system.
A full pass of feedforward and backpropagation is what is referred to as an ``epoch'' and is one of the hyperparameters (parameters that are not learned by the system) to be tuned while training.

The feedforward nature of the network refers to the images being fed through the network from input to output.
Initialisation can be crucial to the performance of an NN.
A random or zero intialisation of the weights can lead to the network taking longer than needed to learn and so to reduce the number of epochs needed we employ what is known as He initialisation \citep{2015He2}.
This initialises the weight matrix to being randomly sampled from a Gaussian distribution $\mathcal{N}$(0,$\sigma$) where
\begin{equation}
    \label{Eq-init}
    \sigma = \sqrt{\frac{2}{n_{l}}},
\end{equation}
where $n_{l}$ is the number of connections in layer $l$ (\textit{i.e.} the number of feature maps per convolution layer in a CNN).
This result is derived from taking the variance of the forward linear process in the neural network.
This initialisation led to the first machine learning algorithm that out-performed a human in image classification \citep[see][for more details]{2015He2}.

At the end of the feedforward process, the network uses the current weights to calculate what class the image belongs to in its current understanding.
Then backpropagation begins: the gradient of the distance between the obtained class label and the true class label is calculated and then fed backwards through the network updating each weight as it goes such that next time this process begins the number of incorrect classifications will be reduced (assuming that our optimiser has taken a step in the right direction in our loss space).
The optimisation technique used for backpropagation is stochastic gradient descent (SGD).
This parameterises the distance between what the network thinks and the truth by an analytical function known as the loss function.
This is a first-order gradient method and the updates to the weights can be calculated very simply mathematically:
\begin{equation}
    \label{Eq-sgd}
    \theta_{t + 1} = \theta_{t} + \eta \nabla_{\theta} L (x; ~\theta_{t}),
\end{equation}
where $\theta_{t+1}$ is the updated weight.
$\eta$ is known as the learning rate which determines how much the correction of the weight will move it throughout the loss space spanned by the loss function (\textit{i.e.} we are optimising the weights on some high dimensional space and the learning rate helps define our walk through that space in search of a minimum).
This is the second of two hyperparameters to be experimented with during training.
The optimisation method we use is SGD with Nesterov momentum \citep{2013Sutskever} which is similar to standard SGD but has a velocity term associated with it leading to an acceleration in the weight updates over many epochs.
Equation~\ref{Eq-sgd} can then be modified to include this velocity:
\begin{equation}
    \label{Eq-sgdnm}
    \theta_{t+1} = \theta_{t} + v_{t+1} = \theta_{t} + \mu v_{t} - \eta \nabla_{\theta} L(x; ~\theta_{t} + \mu v_{t}),
\end{equation}
so rather than $\theta_{t+1}$ being solely updated by the gradient, it is also updated by the product $\mu v_{t}$ where $\mu$ is the momentum coefficient and $v_{t}$ is the velocity for the previous epoch.
The term in the argument of the gradient allows this method to correct the velocity term in a faster way if the prediction we are currently at is not good.
For example, if the product $\mu v_{t}$ results in a poor update for the weight then the gradient function calculated will be steeper and thus tend back towards $\theta_{t}$ such that the optimiser can try again in another direction.
Thus SGD with Nesterov momentum allows us to traverse the loss space at an accelerated rate but, by construction, since areas with flatter curvature will be closer to the minima we are trying to achieve, the acceleration will slow as we approach this minimum and thus we will not overshoot.

Another crucial part of training is hyperparameter tuning and in our network we have three hyperparameters: the learning rate, the number of epochs and the momentum coefficient.
However, we keep the momentum coefficient constant at 0.9 such that the momentum will have a noticeable effect on the gradients and we only have two variable hyperparameters.
The other two hyperparameters are changed during training and a set of models is trained (see Section~\ref{S-results}).
The number of epochs required for general convergence varies from problem to problem.
If the number of epochs is too low then the model will be underfitted and the results cannot be trusted.
If the number of epochs is too high then the model will overfit the data which can lead to misclassifications of unseen data since the network memorises the data structure.
Finding the optimal number of epochs can avoid underfitting but a further measure needs to be taken to avoid overfitting.

A further measure is to use some of the training data not in the training phase but rather as a validation phase.
This means that the network's response to unseen data can be monitored since the validation data will have a defined class that is known beforehand.

Likewise, tuning the learning rate is a problem-specific task.
If the learning rate is too high, it is possible that the system will always skip over minima and never converge to a good solution whereas if it is too low, the system may never escape from a bad local minimum and may converge to a bad solution.

Due to the aforementioned reasons, the process of training a machine learning algorithm must be carefully considered.
This can be difficult, with many tests needed to obtain the desired result.
This is discussed further in Section~\ref{S-results}.

\section{Our Model} \label{S-model}
\begin{figure*}
    \centering
    \includegraphics[width=\textwidth]{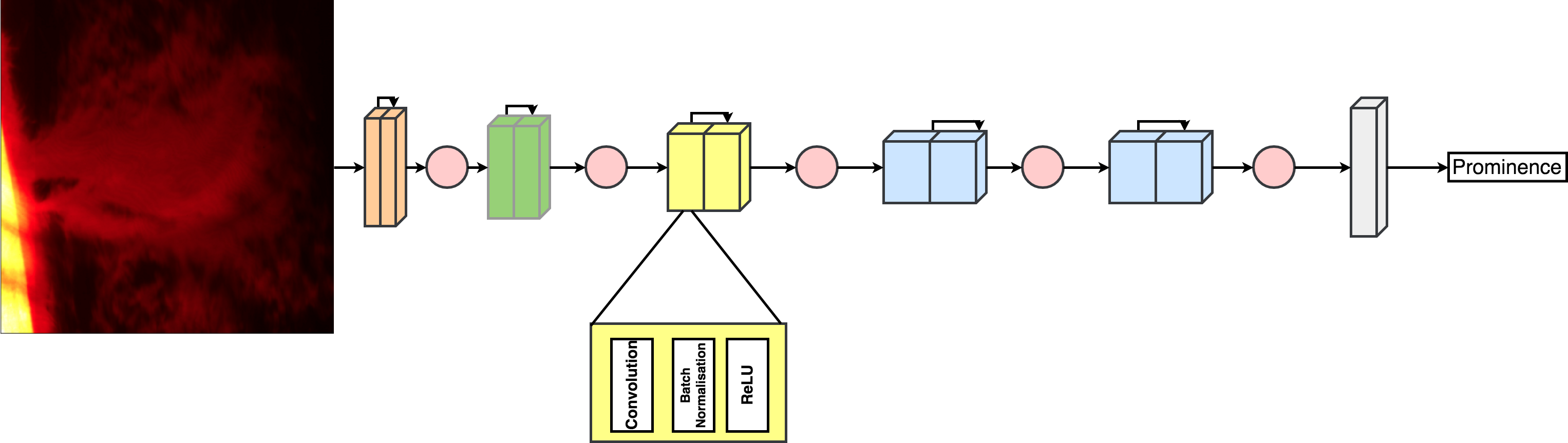}
    \caption{The setup of our 13 layer CNN inspired by VGG networks \citep{2014Simonyan} where the arrows between each block indicate the flow of data in the feedforward process.
    The blocks are colour-coded to reflect their purpose.
    Orange, green, yellow and blue are all convolutional layers which have 64, 128, 256 and 512 trainable feature maps respectively.
    The inside of one of the convolutional layers is shown which is the same for all convolutional layers -- the data undergoes a convolution followed by batch normalisation followed by the activation \textit{via} a ReLU function.
    The red circles correspond to the max pooling layers.
    The grey block corresponds to the classifier at the end of the network.
    The example here is of a prominence in H$\alpha$ from \textit{Hinode}/SOT being classified correctly.}
    \label{F-cnn}
\end{figure*}

The model we introduce is a 13 layer CNN (see Figure~\ref{F-cnn}).
The network is looking to model the function that maps an image of the Sun in H$\alpha$ to a vector of probabilities of the images containing a specific feature.
Therefore, the input of the network will be the pixel intensities of the image and the output will be a vector of class probabilities with each element corresponding to the probability of a feature.
If the network learns the features correctly then the highest probability (\textit{i.e.} the maximum value element of this vector) will correspond to the correct class for the image.

The layers shown as cuboids in Figure~\ref{F-cnn} between the input and the output are essentially doing many, large matrix computations wherein the convolution of the weight matrix and the input to a layer is found; then the batch normalisation is performed on the results of these calculations and, finally, the activation (non-linearity) function is applied to the result to determine the signal passed to the next layer.
The convolution kernels in each of these layers is composed of 3$\times$3 pixels initialised by He initialisation described in Section~\ref{S-training}.
The values of the convolution kernels are the learnable parameters in this model for these layers \textit{i.e.} the values of the kernels are updated during training by the optimiser such that the network learns what abstract features being picked out by convolutions correspond to specific physical features.
The number of feature maps in each of these layers increases towards the output of the network as the model is detecting more and more complex features and a larger number of convolutions to look at will help to distinguish between these features.

``Batch normalisation'' \citep{2015Ioffe} is applied to the output from the convolution operation.
This is a technique used to increase the stability of our network and normalises the output of the convolution calculation around a batch mean ($\beta$) and standard deviation ($\gamma$) \textit{via} the equation:
\begin{equation}
    \label{Eq-bn}
    y = \gamma \times \frac{x - E[x]}{\sqrt{\sigma(x) + \epsilon}} + \beta,
\end{equation}
where $x$ is the output feature maps and $y$ is the batch normalised feature maps, $\epsilon$ is a small positive constant used to stop the denominator going to zero and $\sigma$ is the sample variance of the feature maps being normalised.
This is beneficial as it reduces the dynamic range of the data at the cost of two extra trainable parameters ($\beta$,$\gamma$) and speeds up training sufficiently (if the batch size is large enough).
Equation~\ref{Eq-bn} can then be easily manipulated during backpropagation to return $x$ such that the true feature maps can be recovered from the batch normalised feature maps.

After batch normalisation, the data undergoes a non-linear transformation known as an ``activation function''.
This is a function which shifts the output of the batch normalisation onto a different distribution which determines the signal being passed onto the next layer of the network.
For this function, we use the rectified linear unit \citep[ReLU;][]{2010Nair} function:
\begin{equation}
    \label{Eq-relu}
    \phi (x) = \mathrm{max} (0, x).
\end{equation}
This is chosen due to the sparsity of the output increasing training speed and its ability to avoid the vanishing gradient problem.
The vanishing gradient problem is when the gradients of the loss function during backpropagation becomes so small that they tend to zero and so the optimiser gets stuck in the loss space.
This is avoided when using ReLUs since the gradients of these will never be small:
\begin{equation}
    \label{Eq-relud}
    \frac{\mathrm{d} \phi}{\mathrm{d} x} = H (x),
\end{equation}
where $H (x)$ is the Heaviside function.
However, ReLUs can get stuck if the batch normalised data is all negative but the network should learn that the batch normalisation parameters should not shift the data into a distribution where it is all negative.

Between the deep layers of the network, there are occasionally maxpooling layers (shown by the red circles in Figure~\ref{F-cnn}.
This is used as a downsampling of the data to increase computational efficiency by reducing the number of parameters, and since this results in less spatial information about the features this will reduce over-fitting and increase translational invariance due to the reduction of the pixel-location-specific data.
This downsampling works by parsing the image into segments of 4 pixels (2$\times$2 grid) and taking the maximum of those pixels.
This means that one pixel in a downsampled image is representative of the four pixel block it came from.
This is, in a sense, how the network learns more complex features -- as the resolution of the input is decreased, each pixel represents more information from the original input and thus each operation is performed on a larger fraction of the original image (\textit{e.g.} 4 pixels rather than 1) which will highlight more complex, larger features \textit{via} the convolution operation.
Other types of pooling exist, such as average pooling (taking the average of the group of pixels we are downsampling), but maxpooling is the prevailing due to its benefits for reducing over-fitting since the same pixel out of the four may not be the maxiumum after every weight update.

\begin{figure}
    \centering
    \includegraphics[width=0.49\textwidth]{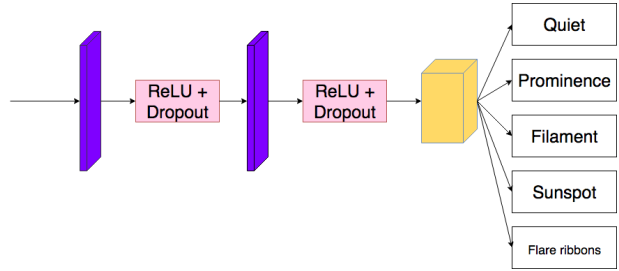}
    \caption{The classifier mini-network.
    The 3D blocks represent fully-connected layers which map the output feature maps from the last maxpooling layer to the class labels with a certain probability.
    The pink boxes refer to rectified linear unit (ReLU) activation followed by dropout regularisation.}
    \label{F-classifier}
\end{figure}

The grey cuboid at the end of the network in Figure~\ref{F-cnn} is the classifier of our network: after the features within the images are identified by the convolutional layers, they are passed to the classifier which decides what class to assign to the images.
This can be described by a mini-network shown in Figure~\ref{F-classifier}.

The two key concepts of this mini-network are the fully-connected block and the dropout regularisation.
The fully-connected layer maps all of the inputs to all of the outputs \textit{via} a linear transformation which results in \textit{N}$\times$\textit{M} parameters to be changed for a layer with \textit{N} input nodes and \textit{M} output nodes.
Dropout is a newer concept in machine learning \citep{2014Srivastava}.
This assigns a probability, \textit{p}, to each input node in a layer such that for each training epoch there is a probability that the network will ignore that node and connection and thus train on an approximate model.
Training on a set of approximate models and then averaging them at validation time works well as a regularisation technique -- \textit{i.e.} helps reduce over-fitting -- whilst still preserving (and actually improving, in many cases) results as shown in \citet{2014Srivastava}.
In our model, we set \textit{p}~=~0.5 (\textit{i.e.} 50\% chance of the node being dropped).
The third fully-connected layer (in gold in Figure~\ref{F-classifier}) determines which class each image should be assigned.
Normally, we would have a final activation function here to do so but the class labels are inferred in our model \textit{via} our choice of loss function which implicitly adds this final activation layer (see Section~\ref{S-tom}).

\subsection{Training our Model} \label{S-tom}
The previous paragraphs have described the feedforward part of our network \textit{i.e.} the path the images take through the network.
We will now described the backpropagation that takes place wherein automatic differentiation \citep{2014GunesBaydin} is carried out on the loss function at each layer to update the weights of the network and learn the classification function optimally.
The loss function we choose to minimise is known as the ``Cross Entropy Loss'' (CEL).
This is based on the discrete version of multinomial logistic regression
\begin{equation}
    \label{Eq-cel}
    L (x) = - \sum_{i} p (x_{i}) \log q (x_{i}),
\end{equation}
where $x$ is the output vector of the network, $p (x_{i})$ is the true class label of the image (\textit{i.e.} $p~=~1$ for the true class label and $p~=~0$ for the other labels) and $q (x_{i})$ is the estimated probability of the label of the image by our model, where we model the classes as being distributed by a softmax distribution
\begin{equation}
    \label{Eq-softmax}
    q (x_{i}) = \frac{\exp(x_{i})}{\sum_{k}~\exp(x_{k})}.
\end{equation}
This softmax function encompasses the last implicit activation (mentioned above) due to the truncating nature of the exponential function (\textit{i.e.} if the network thinks an image is a certain class then $q\to1$).
This loss function is smooth and differentiable and the goal is get $q$ as close to 1 as possible for every image.
In Equation~\ref{Eq-softmax}, $x_{i}$ is then the composite map of all of the layers and activations which can be differentiated \textit{via} the chain rule to make propagating the weight updates between layers as trivial as matrix calculations.

Having described how the network operates, we will now decribe how to train the network.
We train the network from H$\alpha$ ($\lambda$~=~6563~\AA{}) \textit{Hinode}/SOT data.
We have 13175 images in our dataset which have one of the five features in them which were classified by a human (these images are split evenly between the classes to avoid introducing observational bias into the training data).
This is split 90\% to 10\% between training and validation (11857 and 1318, precisely).
We then train over 100 epochs performing the validation after each epoch to find the highest classification percentage.
We perform this for a set of constant learning rates $\eta$~=~\{10$^{-3}$, 5$\times$10$^{-4}$, 10$^{-4}\}$.
This gives us a good idea of which pair of parameters gives the best convergence -- this is a brute-force approach to hyperparameter optimisation but worked well for our model.
This is discussed more in Section~\ref{S-valsot}.
Per learning rate, our model takes $\approx$3 hours to train and validate over the 100 epochs on an NVIDIA Titan Xp.
We train and validate with a batch size of 32 (due to GPU memory limitations).
A higher batch size leads to shorter training time whilst simultaneously improving the batch statistics leading to a higher accuracy.

We train the network to learn the geometry of five solar features: filaments, H$\alpha$ flare ribbons, prominences, sunspots and the quiet Sun (absence of these other four features).
While filaments and prominences are the same physical feature \citep[dense, cool plasma that runs parallel to a magnetic neutral line and is suspended in the atmosphere by a coronal magnetic field;][]{2011Fletcher}, just in different locations (prominences off-limb; filaments on-disk), their geometries in the \textit{Hinode}/SOT H$\alpha$ images are vastly different leading to the split in classification.
This split can easily be consolidated when using the network by asking for images with both filaments and prominences.

The H$\alpha$ flare ribbons are intense brightenings in the solar atmosphere which are interpreted as the base of the coronal magnetic field structures which attribute to flare energisation.
The images of sunspots either contain one of multiple sunspots such that our network learns what a singular sunspot looks like but can still understand if there is a group.
This distinction may become more important and could branch off into two separate classes.
See Section~\ref{S-discussion} for more information.

\section{Results} \label{S-results}
\subsection{Validation on Unseen \textit{Hinode}/SOT Data} \label{S-valsot}

\begin{table}[]
    \begin{tabular}{cllllll}
    \hline
    \multicolumn{1}{l}{} &  & \multicolumn{5}{c}{Network Classification} \\
    \multicolumn{1}{l}{} &  & Filaments & Flares & Prominences & Quiet & Sunspots \\ \hline
    \multirow{5}{*}{True Classification} & Filaments & 175 & 1 & 0 & 0 & 0 \\
     & Flares & 0 & 270 & 0 & 0 & 0 \\
     & Prominences & 0 & 0 & 304 & 0 & 0 \\
     & Quiet & 0 & 0 & 0 & 242 & 0 \\
     & Sunspots & 0 & 0 & 0 & 0 & 326 \\ \hline
    \end{tabular}
    \caption{The confusion matrix for our deep CNN.
    This is a representation of the network's performance on the validation set where each element in the confusion matrix is the number of images classified as containing a feature compared to the true feature contained in that images.}
    \label{T-cm}
\end{table}

\begin{figure*}[t]
    \centering
    \includegraphics[width=\textwidth]{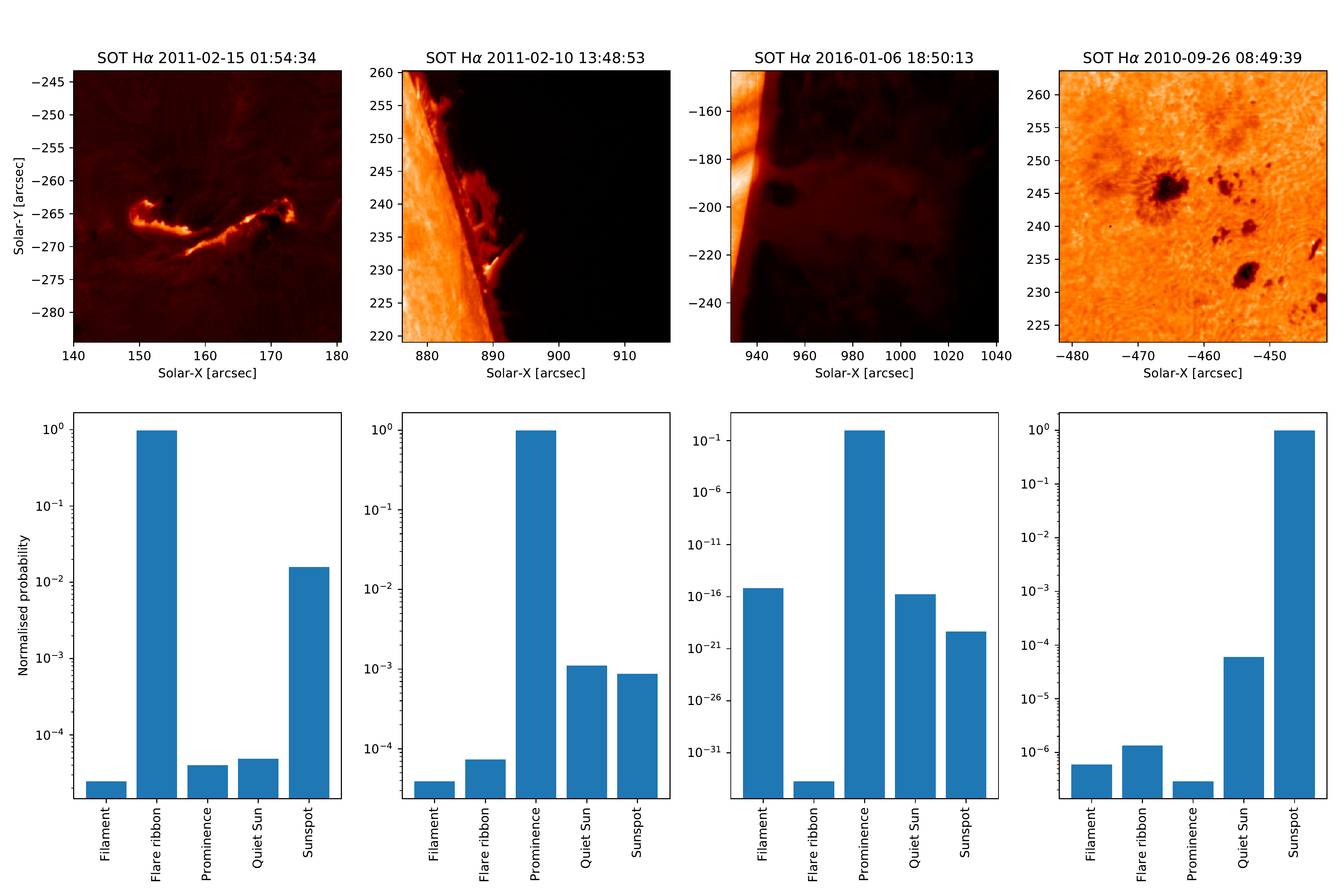}
    \caption{Validation on unseen images from \textit{Hinode}/SOT imaged in H$\alpha$.
    This shows our network correctly identifying images with flare ribbons (left column), prominences (middle columns) and sunspots (right columns) in images it has never seen before.}
    \label{F-val}
\end{figure*}

After training, we test our network on a validation set of images that the network has never seen before taken randomly from the training set.
This validation set consists of 1318 \textit{Hinode}/SOT H$\alpha$ images roughly evenly distributed between the classes.
Validation takes 4.66 seconds on our NVIDIA Titan Xp and, equivalently, 895 seconds on an Intel Core i7-8700 3.20GHz CPU (but could be parallelised over multiple cores to be faster).

The results of training and validating over the ranges of hyperparameters discussed in Section~\ref{S-tom} leads to us learning that our optimal hyperparameters are $\eta$~=~5$\times$10$^{-4}$ and number of epochs of training, \textit{n}~=~5.
This gives a validation accuracy of 99.92\% (1 out of the 1318 images are misclassified\textbf{, see Appendix~\ref{S-va}}).
As the classification percentage is not 100\%, we can conclude that our model has not encapsulated the entirety of the space containing the function which maps the input of our network to the output \textit{i.e.} our minimised loss function is not the optimally minimised loss function, but has learned enough of this space to generalise to unseen data.
The near-perfection of this model is impressive and not to be understated, as a perfect classifier for image data is difficult to come by due to the possibility of distortions and artifacts leading to misclassification.
A further, deeper exploration of the hyperparameter space may lead to even better classification since we have only chosen discrete steps in this space.

Classification percentage on a validation set is however not statistically robust enough to determine whether or not our classifier has actually learned what we wanted it to.
This can be a result of having an uneven split in the validation set between the classes or having a strongly biased classification task.
To deal with this, we calculate the ``confusion matrix'' for our classifier.
This is a matrix whose elements correspond to what class an image actually belongs to compared to what class the network classified it in.
This is shown in Table~\ref{T-cm}.
This tells us about different kinds of errors our network makes.
The predictions our network made can now be split into 4 categories for each features:
\begin{enumerate}[label=\roman*)]
    \item True positives: the number of images containing the feature we are interested in that are correctly identified as containing that feature.
    That is, for a feature $i$ that is of interest to us:
    \begin{equation}
        \label{Eq-tp}
        \mathrm{tp}_{i} = c_{ii},
    \end{equation}
    where $c_{ij}$ is an element of the confusion matrix.
    \item False positives: the number of images not containing the feature we are interested in that are identified as containing that feature
    \begin{equation}
        \label{Eq-fp}
        \mathrm{fp}_{i} = \sum_{i = 1}^{n_{\mathrm{rows}}} c_{ki} - \mathrm{tp}_{i}.
    \end{equation}
    \item False negatives: the number of images containing the feature $i$ that are misclassified as not containing the feature.
    \begin{equation}
        \label{Eq-fn}
        \mathrm{fn}_{i} = \sum_{l = 1}^{n_{\mathrm{cols}}} c_{il} - \mathrm{tp}_{i}.
    \end{equation}
    \item True negatives: the number of images not containing feature $i$ that are correctly classified as not containing feature $i$
    \begin{equation}
        \label{Eq-tn}
        \mathrm{tn}_{i} = \sum_{k = 1}^{n_{\mathrm{rows}}} \sum_{l = 1}^{n_{\mathrm{cols}}} c_{kl} - \mathrm{tp}_{i} - \mathrm{fp}_{i} - \mathrm{fn}_{i}.
    \end{equation}
\end{enumerate}
From these measures we can define two statistics that can probe how well our classifier works.
The first is known as ``precision'' and this tells us how many images that our model classified as having feature $i$ truly contain feature $i$
\begin{equation}
    \label{Eq-prec}
    \rho_{i} = \frac{\mathrm{tp}_{i}}{\mathrm{tp}_{i} + \mathrm{fp}_{i}}.
\end{equation}
The second is known as ``recall''.
This tells us about the percentage of images containing feature $i$ that were correctly identified as containing feature $i$.
This can be thought of as the ability of the model to find all of the images of interest to us
\begin{equation}
    \label{Eq-re}
    r_{i} = \frac{\mathrm{tp}_{i}}{\mathrm{tp}_{i} + \mathrm{fn}_{i}}.
\end{equation}
Ideally we want precision and recall to both be equal to one for all classes.
The precision for flare ribbons deviates from one as the misclassified image is misclassified as a flare ribbon.
This corresponds to the image not containing a flare ribbon but the network thinking it does.
The precision for all other classes is one meaning that the network does not think any images not containing these features actually contain these features.
The recall for filaments is the only recall different from one as it is an image containing a filament that is misclassified.
This means that the network thinks this image containing a filament actually contains another feature (in this case a flare ribbon).
The recall being equal to unity for all other classes means that the network never classifies any of those images as having a feature different to the feature they contain.
Overall, the misinterpretation of our network is not detrimental to its performance.
We are confident that our network has learned the geometry of these features due to the tiny margin of error it has.

Figure~\ref{F-val} shows examples of our network classifying images.
These are images from SOT in H$\alpha$ which the network has not seen during training.
This provides a test to ensure our network is not ``memorising'' the training data \textit{i.e.} adjusting its weights to classify only the training set correctly.\footnote{We have also tested our network on images in other visible wavelengths, an example of which can be found at \url{http://github.com/rhero12/Slic/blob/master/testing_example.ipynb}.}

The left column of Figure~\ref{F-val} shows an image with clear flare ribbons that are classified correctly by the network.
However, there is also a sunspot in this image which the network picks up on in the probability distribution (left column, bottom).
There is a non-negligible probability that the important feature in this image is a sunspot.
This means that we can use our network for classification of multiple large scale features in a single image.
However, a more precise way to do multi-label classification would be more beneficial and is discussed in Section~\ref{S-discussion}.
The other images are classified correctly in Figure~\ref{F-val} showing that our model has learned the geometry of these features.

\subsection{Testing on Other Instruments' Imaging Data} \label{S-testaia}
Having trained the network on \textit{Hinode}/SOT H$\alpha$ data, we performed adversarial tests on the network.
These are tests in which the input to the network is designed to be confusing to the network.
We focus on adversarial examples where the answer is obvious to the user but not necessarily to the network.
That is, we perform tests on the network for sunspots and prominences in different wavelengths given that they look perceptually similar to the features in H$\alpha$.
To do this, we use data from UV wavelengths for sunspots and EUV wavelengths for prominences from the SDO/AIA instrument \citep{2006Title}.
This gives an idea of whether or not we can generalise our network to other wavelengths without retraining.

\subsubsection{Sunspots in UV} \label{S-ss}
\begin{figure*}[t]
    \centering
    \includegraphics[width=\textwidth]{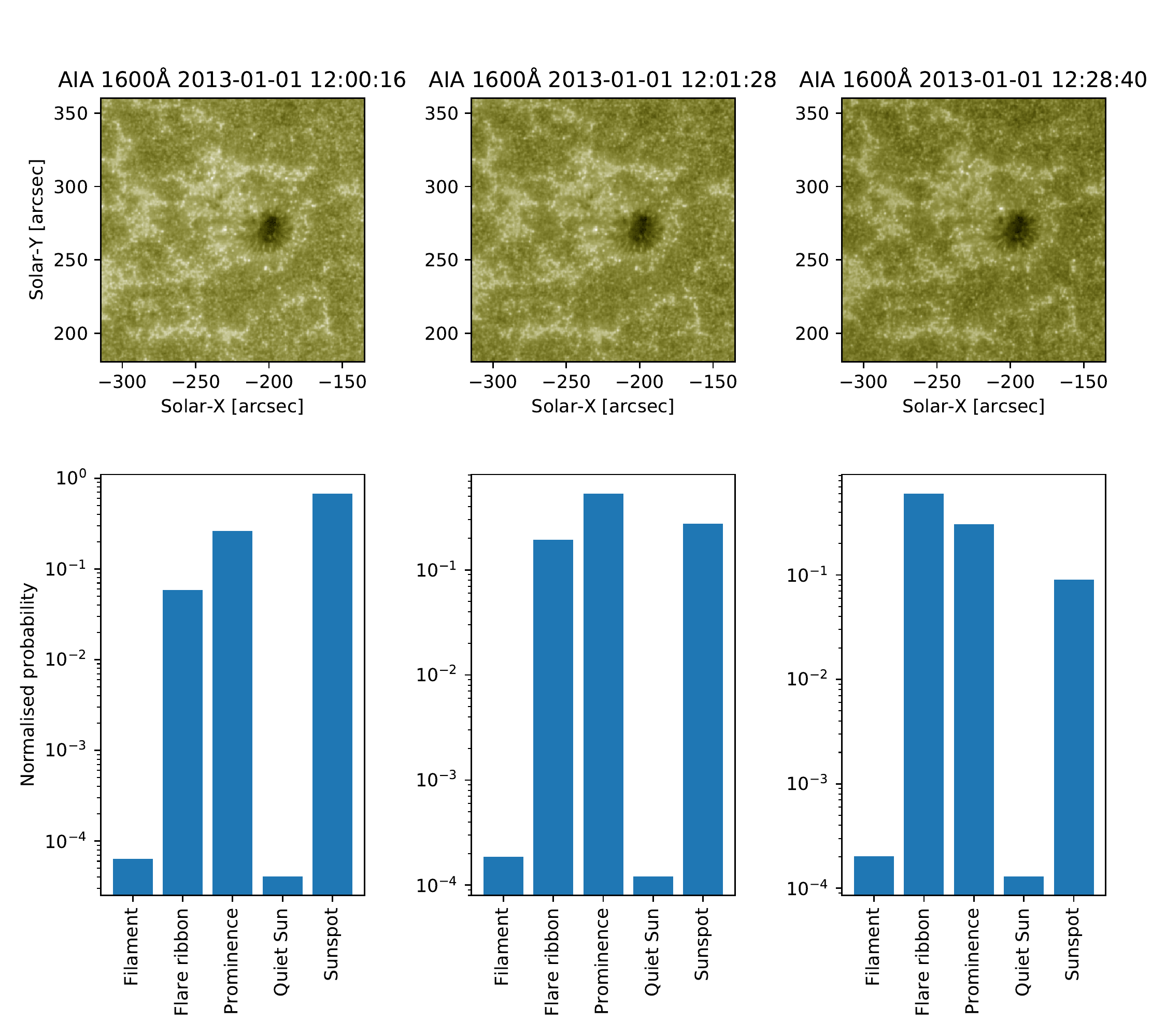}
    \caption{The three images analysed here are of the same sunspot (AR11638) imaged in SDO/AIA 1600\AA{} a few minutes apart.
    These are shown to highlight the confusion of the network when dealing with UV sunspots as sometimes the sunspot is classified correctly while other times it is not.
    The sunspots in UV are always either classified as sunspots, flare ribbons or prominences.
    We hypothesise this to be due to the elongated, bright plage in the sunspot images we test on.
    We found this to be true also for AR12665 and AR12674.}
    \label{F-ss1}
\end{figure*}
We use three different sunspot datasets observed in both 1600 \& 1700\AA{} (the AIA UV channels): AR11638 from 2013/01/01, AR12665 from 2017/07/10 and AR12674 from 2017/09/06.
Each dataset used in our examples is over a 1 hour time range (12:00:00-13:00:00UTC).

The UV sunspot data was not classified well by our network.
Despite the sunspots in each active region not evolving much over the observed time range, the network sometimes classifies these sunspots correctly whilst sometimes incorrectly classifying them as either flare ribbons or prominences.
An example of this for AR11638 imaged in 1600\AA{} is shown in Figure~\ref{F-ss1}.
We hypothesise two possible reasons for this:
\begin{enumerate}[label=\roman*)]
    \item There are small-scale UV brightenings around sunspots.
    This can be attributed to plage (dispersed brightenings in an active region).
    While these brightenings occur in the optical and the ultraviolet; they are more noticeable in the UV due to the background UV quiet Sun being dimmer than in the optical.
    This implies that the contrast between plage and quiet Sun in UV wavelengths will be higher which can impact the network's classification ability by convincing it that the brightenings are the important feature.
    Furthermore, the plage can often look like elongated bright regions and this elongation may be further proof to the network that this image should be classified as something other than a sunspot.
    \item The lack of spatial resolution in the AIA images.
    In H$\alpha$, SOT has a spatial resolution of 0.33$^{\prime \prime}$ \citep{2008Tsuneta} whereas for the AIA UV channels the spatial resolution is 1.2$^{\prime \prime}$ \citep{2006Title}.
    This disparity could be another cause (or composite cause) of the misclassification of the UV sunspot observations.
    Due to the nature of convolutional feature extraction, the extracted features from two images of the same object but with different resolutions can be vastly different.
    This would affect the feature maps being passed through our CNN and thus the end classification result.
\end{enumerate}

We test the first hypothesis using these three datasets and the results are illustrated in Figure~\ref{F-ss1} (and in Figures~\ref{F-app1} and \ref{F-app2}).
Due to the incorrect classifications being either flare ribbons or prominences for these active regions we believe that the brightness and elongation of the plage region is responsible for this.
As can be seen in Figure~\ref{F-val}, in H$\alpha$ both flare ribbons are bright, elongated structures on a darker background which is what leads us to believe the first hypothesis is responsible for incorrect classification.

\begin{figure*}[ht]
    \centering
    \includegraphics[width=\textwidth]{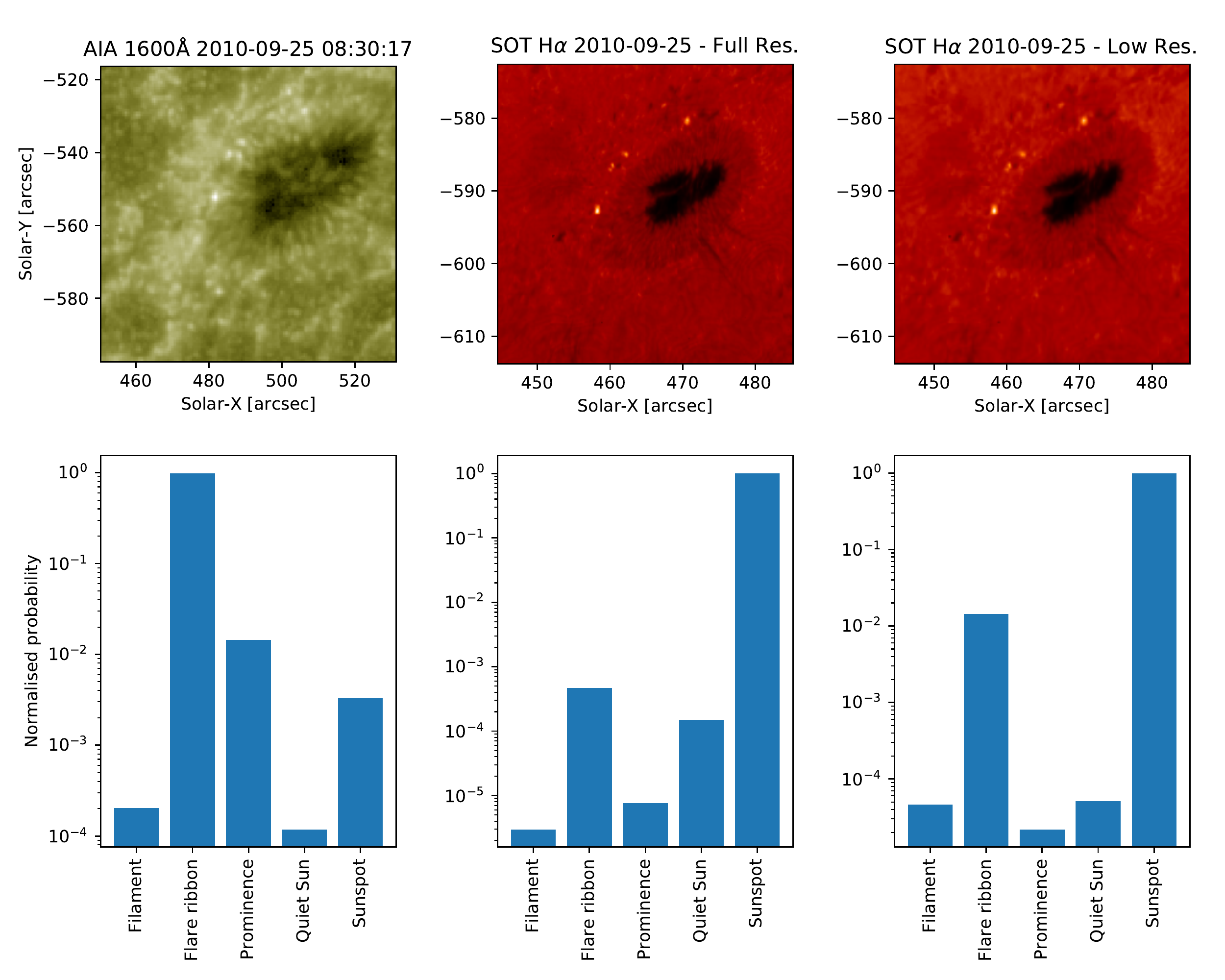}
    \caption{Comparison of the sunspot from AR11108 images in SDO/AIA 1600\AA{} (left column) and \textit{Hinode}/SOT H$\alpha$ at full resolution and degraded to AIA resolution (middle and right columns, respectively).
    This illustrates that the resolution does not play a significant role in skewing the classification of our network as both the full resolution and low resolution SOT images are classified correctly whilst the AIA image is not.}
    \label{F-ss2}
\end{figure*}

To test for the second hypothesis, we must look for sunspot observations that are co-temporal with observations from SOT in H$\alpha$.
The results of this test are shown in Figure~\ref{F-ss2}.
The dataset we chose was from a single-sunspot active region AR11108 from 2010/09/25.
The observations used from AIA were taken from 08:05:00-09:20:00UTC.
An example is shown in the left column of Figure~\ref{F-ss2} where the sunspot was misclassified as a flare ribbon.
The observations used from SOT were taken synoptically with AIA in H$\alpha$.
These H$\alpha$ images are downsampled by $\approx$3$\times$ to AIA resolution.
The full resolution and low resolution images are then passed to the network.
Both sets of images are classified perfectly by the network as shown by the middle and right columns of Figure~\ref{F-ss2}.
This result invalidates the second hypothesis and leads us to the conclusion that resolution is not a determining factor in misclassifications.
Thus, we conclude that the plage is the feature confusing our network from understanding sunspots in UV.

\subsubsection{Prominences in EUV} \label{S-prom}
\begin{figure*}[ht]
    \centering
    \includegraphics[width=\textwidth]{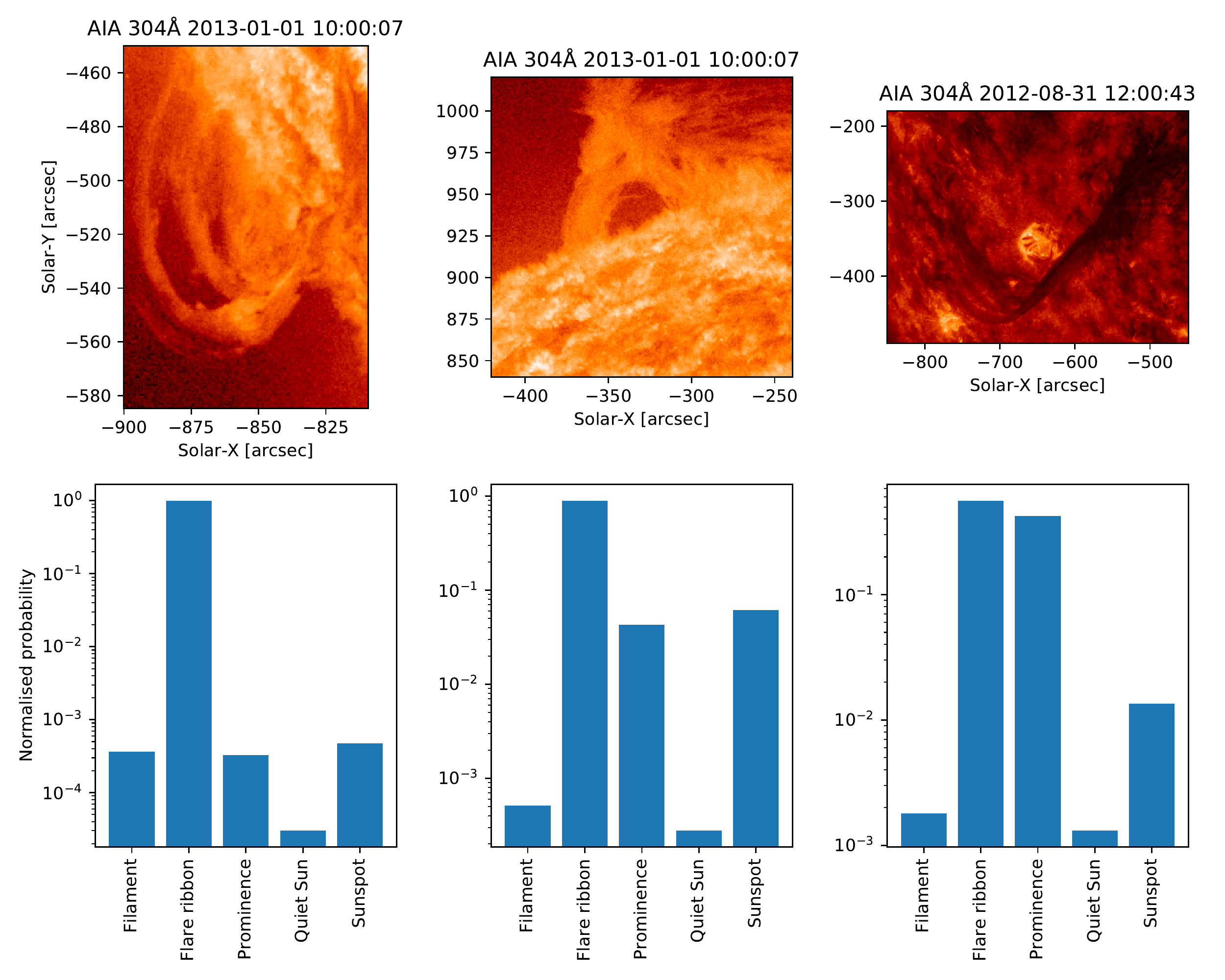}
    \caption{Examples of incorrect classification of prominences obsered in SDO/AIA 304\AA{}.
    The left and middle columns are quite confidently classified as flare ribbons which we hypothesise to be due to the background coronal He\textsc{ii} emission visible in these images but does not have an analogue in the H$\alpha$ training set.
    The right column shows the network thinking that the image is nearly as likely to contain a prominence as a flare ribbon.
    We assume that the network identifies the bright patch in the middle as a flare ribbon but also picks up the prominence above it.
    This shows the network's capability of giving a good idea if there are multiple features in a single image without being explicitly taught to do so.}
    \label{F-prom}
\end{figure*}
We use two different datasets for three different prominences.
We use SDO/AIA 304\AA{} observations to look at prominences which correspond to He~\textsc{ii} emission at $\approx$50,000K.
These datasets were taken from 2012/08/31 12:00:00-13:00:00UTC and 2013/01/01 10:00:00-11:00:00UTC.
The 2012/08/31 dataset has the prominence located off the eastern limb of the Sun and is shown in the right column, top row of Figure~\ref{F-prom}.
The 2013/01/01 dataset has two prominences: one located off the eastern limb north-east from disk centre and another located off the eastern limb south-east from disk centre.
These are shown in the left and middle columns, top row of Figure~\ref{F-prom}.

As shown in the bottom row of Figure~\ref{F-prom}, none of our prominences are predicted correctly.
We hypothesise this to be due to the noisy coronal background emission at the heights of the prominence that we observe.
We see that in the images in Figure~\ref{F-prom}, there is emission in the region of the prominence that is not directly the prominence.
In contrast, the H$\alpha$ images from \textit{Hinode}/SOT do not have emission except in the prominence at the heights of the prominence (as can be seen in the second and third rows of Figure~\ref{F-val}).

All of the images in Figure~\ref{F-prom} are misclassified as flare ribbons.
For the two prominences from 2013/01/01, we hypothesise this to be caused by the background He~\textsc{ii} emission as this causes the prominence to appear bright against an emitting background which is similar to the flare ribbon images used for training in H$\alpha$.
The prominence from 2012/08/31 has comparable probabilities of the image containing a flare ribbon or a prominence.
In the image of this prominence, we assume from the flare ribbon classification that our model chooses the bright point in the middle of the image as the most important feature.
Interestingly though, our network picks up the geometry of the prominence as a different feature and is almost equally confident that this image contains a prominence.
We also make the same argument as in Section~\ref{S-ss} that the difference in resolution does not impact our classification ability.
Therefore, we reach the conclusion that only the coronal emission effects the classification ability of our network.

\section{Discussion} \label{S-discussion}
We have shown that a deep convolutional neural network can learn the geometry of features on the Sun.
This works very well for the wavelength that the network is trained on but does not always generalise to other wavelengths (which is to be expected due to some emission mechanisms occurring in some wavelengths and not others).
This leads to a discussion about how the network can be improved through more detailed classification and multi-wavelength training regimes that could produce a classifier that generalises better to unseen data.
Also increasing the depth of training can lead to more efficient uses of transfer learning from one network to another.

Further improvements to the network will make it more versatile and precise.
In the versatility direction, we plan on devising a multi-label classification.
This means that each image will have more than one label \textit{e.g.} \textit{n} sunspots or single flare ribbon; we plan to analyse this sequentially.
One way to do this is by using multiple binary classifier CNNs on the images and using the results from the binary networks to determine what features are in an image, \textit{e.g.} one network to detect sunspots, one to detect flare ribbons and so on \citep{2011Read}.
Another is using an ensemble method where there is a set of multiclass classifiers that each assign one label to the image.
These predictions are then combined with each class getting a certain percentage of a vote from each classifier and the labels with a percentage above a certain threshold are used as the multi-label for the image \citep{2013Rokach}.
We plan on doing this using a recursive neural network (RNN).
An RNN is a network that is specialised at processing sequential data.
RNNs do this by using the previous layer's output as the dependency for the current layer's input -- \textit{i.e.} there is some function that connects the output of the previous layer to the input of the current layer in a specified sequence (a ``recurrence relation'').
Following \citet{2016Bui}, we would utilise a convolutional RNN (C-RNN) which takes the feature maps from the last convolutional layer in our original network (after activation) as an input and outputs a compact representation of each feature over many convolutions.
This allows the C-RNN to learn a general form for our features (\textit{i.e.} over many convolutional filters).
For multi-label classification, the C-RNN architecture network will generate N RNNs to describe each image by N labels.
For example, if we take an image that contains at least one sunspot and we want to know if it has a single sunspot or multiple sunspots then we will use 2 RNN blocks -- one to predict that the image contains a sunspot and the second to predict how many sunspots are in the image.
This has seen great success in other image classification cases \citep{2016Bui,2016Wang} and could work well for solar images.

There are many changes we can implement that may improve precision.
We could replace the dropout layers with max-dropout or stochastic dropout proposed in \citet{2017Sungheon} which has improved performance on standard datasets.
Another possibility is to change the convolution blocks to residual blocks \citep{2015He} wherein the network learns the residual of a function (the difference between the function and the input) rather than the function itself.
This has been shown to improve speed, performance and how deep a network can be before suffering from the vanishing gradient problem -- when the calculated gradients are close to zero, causing the optimiser to get stuck in the parameter space which renders any further training useless.
This is solved in residual networks \textit{via} the introduction of skip connections which carry the input to the end of a residual block.
This allows the input to travel deeper in the network without being diminished by the layers.

Another interesting property of our network is that it is based on a series of very successful deep CNNs known as VGG networks which were made to learn the ImageNet dataset \citep{2014Simonyan,2009Deng}.
We found that these deep architectures are necessary for solar image classification as shallower networks did not yield sufficient results (even for a simple task such as image classification).
The ImageNet dataset is a well-known database of millions of images that has been classified into thousands of classes.
This has been an incredibly successful approach and is useful in transfer learning.
The pretrained VGG networks have proven to be extremely useful for transfer learning for real-world images \citep{2017Kupyn,2016Johnson}.
We would like to propose a solar ImageNet (SIN).
SIN would be a huge dataset containing features imaged in different wavelengths from different instruments.
We would then train a classifier to learn what these features look like in different solar contexts.
The classification network we present here can be used as a building block for SIN and acts like a VGG network trained on a subset of ImageNet.

This would make a transfer learning approach to solar machine learning extremely plausible and could lead to increased accuracies in deep learning tasks in solar physics compared to the same networks initialised without transfer learning.
For example, this kind of network would be useful in data pipelines for creating catalogues of data and picking up on observations that were targeted at a specific feature but picked up something else too.
Furthermore, the network we have presented can be used in conjunction with already-existing data pipelines where the data may not have a specific target specified in the meta information.
Due to its speed and accuracy, this model will be useful for anyone having to sift through terabytes of data.
Lastly, networks of this design could be utilised in automating telescope pointing.
With more detailed training (described above), a sufficient network could parse synoptic observations of the observer's target and calculate where the target will be when the observations will be occurring.
The code is available online under the MIT license\footnote{More information available at \url{https://opensource.org/licenses/MIT}.} at \url{https://github.com/rhero12/Slic}.
The release of the code at \url{https://github.com/rhero12/Slic/releases/tags/1.1.1/} contains the pre-trained model and scripts on how to use the code for image classification and transfer learning.
The release also includes the prepared training and validation data explained in Sections~\ref{S-intro},~\ref{S-model},~\ref{S-results}.

\begin{acks}[Acknowledgements]
J.A.A. acknowledges support from `ScotDIST' doctoral training centre supported by grant ST/R504750/1 from the UK's Science and Technology Facilities Council (STFC).
L.F. acknowledges support from STFC grant ST/P000533/1.
Hinode is a Japanese mission developed and launched by ISAS/JAXA, with NAOJ as domestic partner and NASA and STFC (UK) as international partners.
It is operated by these agencies in co-operation with ESA and NSC (Norway).
The AIA data are provided courtesy of NASA/SDO and the AIA science team.
J.A.A. would like to thank C.M.J. Osborne and P.J.A. Sim\~{o}es for helpful discussions.
The authors would also like to thank the reviewer for helpful comments and corrections.
\end{acks}

\begin{acks}[Disclosure of Potential Conflicts of Interest]
    The authors declare that they have no conflicts of interest.
\end{acks}

\begin{appendix}
\section{Misclassification in Validation Set} \label{S-va}
\begin{figure}[H]
    \centering
    \includegraphics[width=0.65\textwidth]{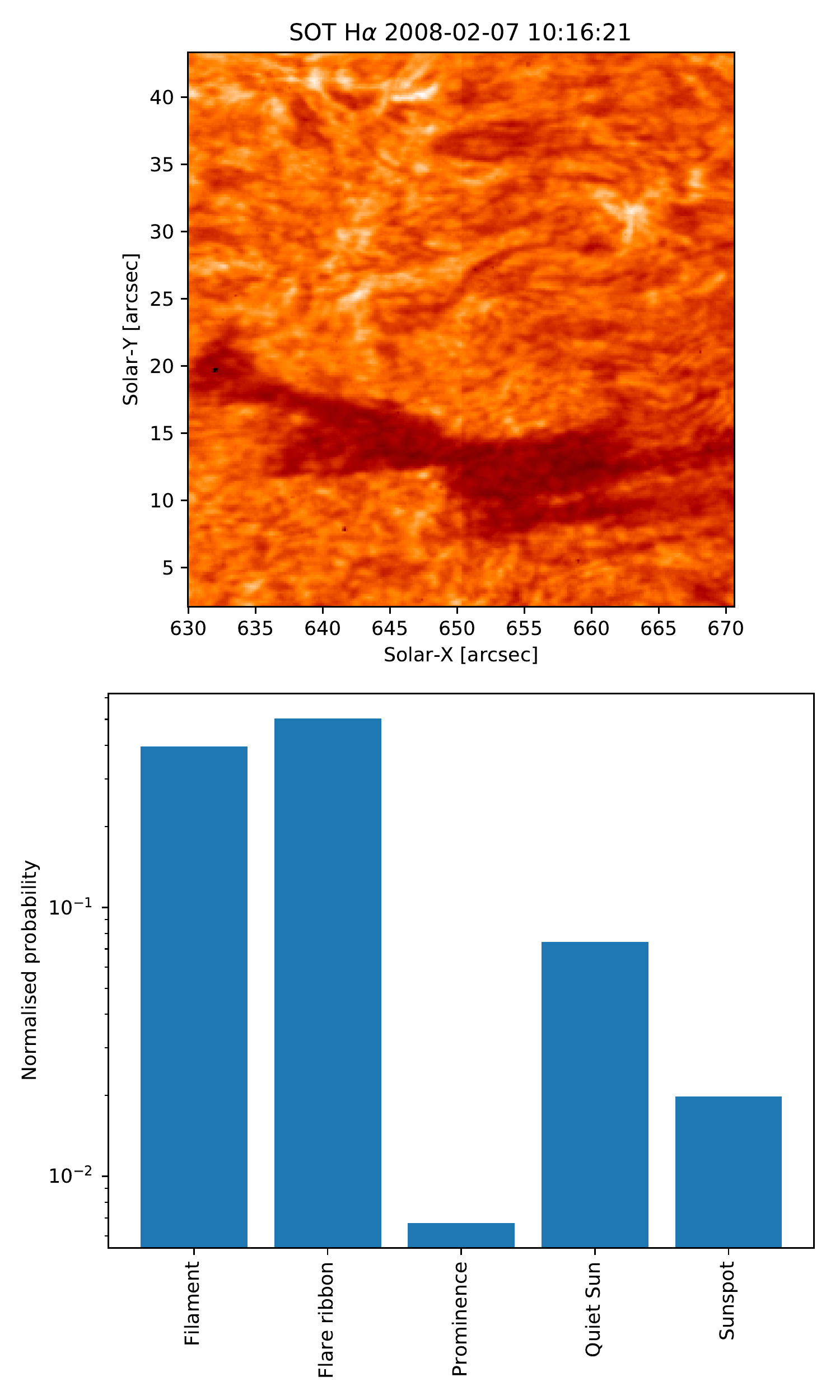}
    \caption{The single misclassified case from our validation set.
    The network identifies the image as containing flare ribbons likely due to the brightenings at the top of the image.
    The image was truly classified by eye as containing a filament.}
    \label{F-vapp1}
\end{figure}

Below we have include the image that was misclassified by our network in the validation stage and its discrete probability distribution showing the likelihood for each class according to our network.
The network classifies the image as harbouring flare ribbons with filament (what the image was actually classified as being a close second).
We believe this to be due to the brightenings in the top portion of the image (see Figure~\ref{F-vapp1}).
However, the probability of filament is non-negligible in comparison and this aids our claim that, as is, our network has the capabilities of detecting multiple features in one image (as discussed in Section~\ref{S-valsot}).
\section{Classification of AR12665 and AR12674} \label{S-ssa}
\begin{figure*}[H]
    \centering
    \includegraphics[width=0.75\textwidth]{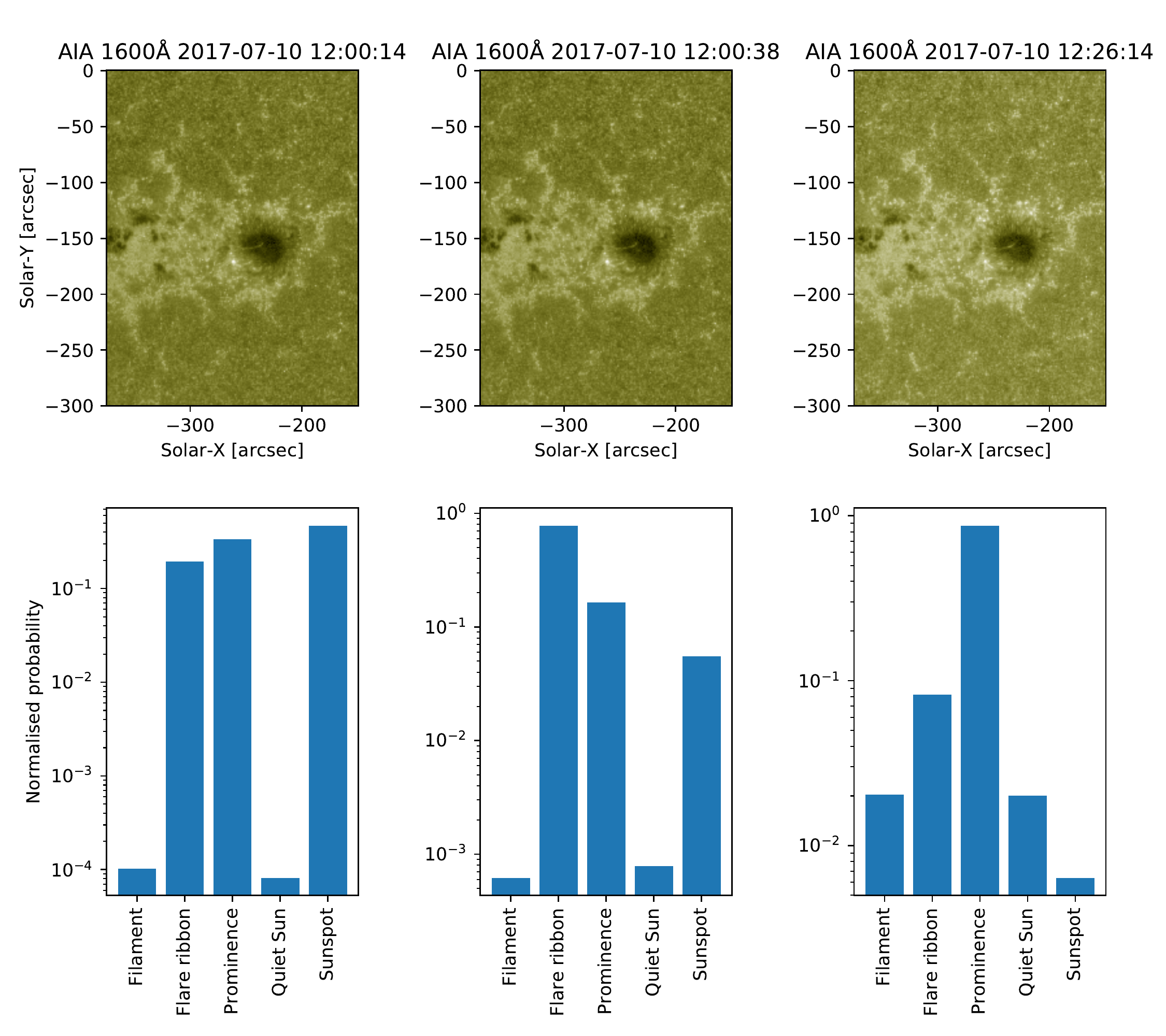}
    \caption{Classification of 1600\AA{} observations of AR12665 showing the same trifecta of different classifications for very similar images as in Figure~\ref{F-ss1}.}
    \label{F-app1}
\end{figure*}

\begin{figure*}[H]
    \centering
    \includegraphics[width=0.75\textwidth]{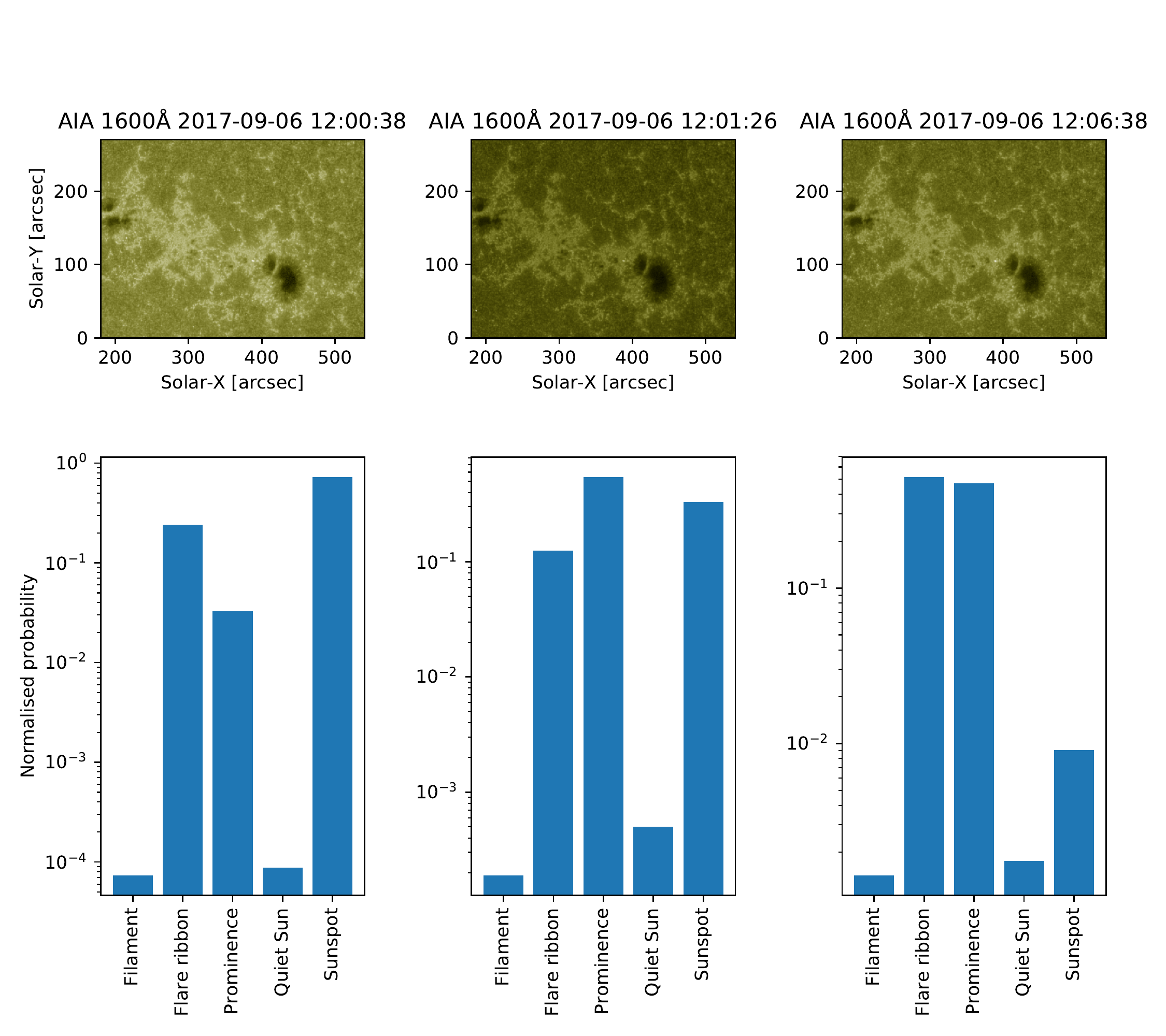}
    \caption{Classification of 1600\AA{} observations of AR12674 similar to Figures~\ref{F-ss1} and \ref{F-app1}.}
    \label{F-app2}
\end{figure*}

We have included the classification of the other two active regions given in Section~\ref{S-ss}.
These are shown in Figures~\ref{F-app1} and \ref{F-app2}.
The classifications of these active regions show the same pattern as the one analysed in the main text as being misclassified as either a flare ribbon or a prominence due to the elongated plage regions whilst sometimes being correctly identified as a sunspot.

\end{appendix}

\bibliographystyle{spr-mp-sola}

\end{article}
\end{document}